\documentclass{an}
\usepackage{graphicx}
\usepackage{times}
\usepackage{fancyhdr}
\begin{document}

\title{A Search for wide visual companions of exoplanet host stars. \newline The Calar Alto Survey}

\author{Markus~Mugrauer\inst{1}, R.~Neuh\"auser\inst{1}, T.~Mazeh\inst{2}, E.~Guenther\inst{3}
M.~Fern\'andez\inst{4}$^{,}$\inst{5} \and C. Broeg \inst{1}} \institute{Astrophysikalisches
Institut und Universit\"ats-Sternwarte Jena, Schillerg\"asschen 2-3, 07745 Jena, Germany\and Tel
Aviv University, 69978 Tel Aviv, Israel \and Th\"uringer Landessternwarte, Sternwarte 5, 07778
Tautenburg, Germany\and Instituto de Astrof\'isica de Andaluc\'ia, CSIC, Apdo.  Correos 3004, 18080
Granada, Spain \and Max-Planck-Institut f\"ur Astronomie, K\"onigstuhl 17, D-69117 Heidelberg,
Germany}

\date{Received; accepted; published online}

\abstract{We have carried out a search for co-moving stellar and substellar companions around 18
exoplanet host stars with the infrared camera MAGIC at the 2.2m Calar Alto telescope, by comparing
our images with images from the all sky surveys 2MASS, POSS I and II. Four stars of the sample
namely HD\,80606, 55\,Cnc, HD\,46375 and  BD$-$10$^{\circ}$3166, are listed as binaries in the
Washington Visual Double Star Catalogue (WDS). The binary nature of HD\,80606, 55\,Cnc, and
HD\,46375 is confirmed with both astrometry as well as photometry, thereby the proper motion of the
companion of HD\,46375 was determined here for the first time. We derived the companion masses as
well as the longterm stability regions for additional companions in these three binary systems. We
can rule out further stellar companions around all stars in the sample with projected separations
between 270\,AU and 2500\,AU, being sensitive to substellar companions with masses down to
$\sim$\,60\,$\rm{M_{Jup}}$ (S/N\,=\,3). Furthermore we present evidence that the two components of
the WDS binary BD$-$10$^{\circ}$3166 are unrelated stars, i.e this system is a visual pair. The
spectrophotometric distance of the primary (a K0 dwarf) is $\sim$\,67\,pc, whereas the presumable
secondary BD$-$10$^{\circ}$3166\,B (a M4 to M5 dwarf) is located at a distance of 13\,pc in the
foreground.}
\correspondence{markus@astro.uni-jena.de}

\maketitle

\section{Introduction}

More than 100 stars are already known to host exoplanets. Detected by radial velocity searches,
theses planets orbit their host stars on relative close orbits with semi-major axes smaller 5\,AU.
Some of these exoplanets were found in binary systems, in which the exoplanet host star is mostly
the more massive primary star. First statistical differences between planets orbiting single stars
and planets located in binary systems were already reported by Zucker \& Mazeh (2002) and
Eggenberger et al. (2004), using samples of 9 and 15 binaries, respectively. In particular, planets
with orbital periods shorter than 40 days exhibit a difference in their mass-period and
eccentricity-period distribution. Mugrauer et al. (2005b) updated and extended the sample of
binaries among the exoplanet host stars (21 binaries) and confirmed the reported statistical
differences for short period planets. These statistical differences between planets orbiting single
stars and planets residing in a binary systems might be an implication of the host star
multiplicity.

Today only four of the binary systems with exoplanets are known to have small projected separations
of the order of 20\,AU (HD\,188753, $\gamma$\,Cep, HD\,41004 and Gl\,86). Due to the proximity of
the stars in these four stellar systems, they are a challenge for planet formation theories (core
accretion and gravitational collapse scenario). In these systems the size of a protoplanetary disk
around the exoplanet host star is truncated by its companion star to only a few AU and does not
extend beyond the ice line (Pichardo et al. 2005). Furthermore, objects revolving around the
exoplanet host star farther outside are perturbed by the secondary star and are not longterm
stable, i.e. finally these objects will be ejected from the system or collide with one of the two
stars (Holman \& Wiegert 1999). Therefore, planets orbiting a component of a close binary system
should be formed and reside only in the adjacency of their parent star. On the other hand, close to
the star, within the ice line, the core accretion (lack of solid material to form the planet core
in a solar minimum mass nebula) as well as the gravitational collapse scenario (high disk
temperature dumps gravitational instability) cannot explain the formation of gas giant planets
(Jang-Condell 2005).

Nevertheless, there are ways around this problem. The secondary star could excite density waves in
the disk, increasing the surface density in some parts of the disk, leading to planet formation via
gravitational instability. Or, the planet bearing disk might be different from the disk of our own
solar system. If disks of the same mass differ only in their angular momentum, such that in smaller
disks more mass is closer in, planets indeed might be formed close to the star within the iceline
(Hatzes \& Wuchterl 2005).

The problem to form planets in close binaries by both the gravitational instability and the core
accretion scenario clearly demonstrates that planets detected in these systems are most intriguing
objects. They provide the possibility to study the effect of stellar multiplicity on the planet
formation, the longterm stability and evolution of planetary orbits and yield ancillary conditions
for the formation of Jovian planets.

So far only few programs have searched for additional visual companions of exoplanet host stars.
Adaptive optics (AO) imaging search campaigns reported several new close companions during the last
years. Patience et al. (2002) detected companions close to HD\,114762 and $\tau$\,Boo. Furthermore
Els et al. (2000) reported a faint companion, separated from the exoplanet host star Gl\,86 by
$\sim$\,2\,arcsec. They concluded from infrared photometry that this companion is a late L or early
T dwarf. Queloz et al. (2000) reported a longterm linear trend in the radial velocity of Gl\,86 and
according to Jahrei\ss~(2001) Gl\,86 turned out to be a highly significant $\Delta\mu$ binary after
combining Hipparcos measurements with ground based astrometric catalogues. Both results indicating
that Gl\,86 has a companion of stellar mass. Mugrauer \& Neuh\"auser (2005a) could finally prove
that the companion, firstly detected by Els et al. (2000), is indeed a white dwarf, the first white
dwarf found close to an exoplanet host star.

Despite being successful, AO imaging surveys cannot find wide companions with separations up to a
few thousand AU because of their small field of view (typically only a few tens of arcsec).
Lowrance et al. (2002) presented a new wide (750\,AU) low-mass stellar companion
($\sim$\,0.2M\,$\rm_{\sun}$) which was detected in the digitized plates of the Palomar Obervatory
Sky Survey, a first example that there could be many more of these objects, all with separations
larger than 100\,AU. The whole sample of exoplanet host stars was not systematically surveyed so
far for these objects, i.e. the multiplicity of exoplanet host stars might be much higher than
derived from AO surveys alone. Therefore, we have started a search program for wide companions of
exoplanet host stars and several new binaries were already detected (see Mugrauer et al. 2004 a,b
and 2005 b).

Most of the exoplanet host stars have large proper motions ($\mu$\,$\sim$\,200\,mas/yr), well known
due to precise measurements of the \textsl{HIPPARCOS} satellite (Perryman et al. 1997). Therefore,
real companions can be identified as co-moving objects by comparing images taken with several years
of epoch difference. Photometry and spectroscopy can then confirm the companionship --- the
measured photometry of the companion must be consistent with an object of the given spectral type
at the distance and age of the exoplanet host star. To be sensitive to low-mass faint substellar
companions, we observed all targets in the near-infrared (H band at 1.6\,$\mu$m), as substellar
companions are several magnitudes brighter compared to the visible spectral range. Furthermore, the
contrast between the hot primary and a substellar companion is smaller in the infrared. Hence,
close companions separated from their primary star by only a few arcsec, can be detected.

At the beginning of 2002 we started our multiplicity study of exoplanet hosts stars in the northern
sky with a first imaging campaign carried out at the Calar Alto observatory (Spain). We selected as
targets all exoplanet host stars, published before 2002 and which are observable with small
airmasses (AM $\le$ 1.5) from Calar Alto (37\,$^{\circ}$ latitude), i.e. in total 44 exoplanet host
stars. The first observations were obtained in February 2002 followed by a further observing run in
July 2002. The third and final run was scheduled for September 2002 but clouded out and no data
could be taken. In total we have observed 18 exoplanet host stars. Most of the remaining 26 stars
were observed in the meantime with either UKIRT\footnote{United Kingdom Infrared Telescope} on
Hawaii and/or NTT\footnote{New Technology Telescope} on La Silla.

\section{The Calar Alto Survey - Results}

Our direct imaging campaign was carried out with the 2.2m telescope at the Calar Alto observatory,
using the near-infrared imager MAGIC. This camera is equipped with a 256x256 HgCdTe-detector with a
pixelscale of 640\,mas in its high resolution mode, i.e. 164x164\,arcsec field of view.

All MAGIC images were astrometrically calibrated using the 2MASS\footnote{2 Micron All Sky Survey}
Point source catalogue (Cutri et al. 2003), yielding the detector pixelscale and the offset in the
position angle (PA). The results of the astrometric calibration for both MAGIC runs are summarized
in Table\,\ref{tab1}.

\begin{table}[htb]
\centering\caption{Astrometric calibration of all observing runs.} \label{tab1}
\begin{tabular}{lcc}\hline
epoch & pixelscale & offset PA\\
& [mas/pixel] & [$^\circ$]\\
\hline
02/02 & 640.2$\pm$3.8 & 0.01$\pm$0.19\\
07/02 & 639.1$\pm$3.8 & -0.21$\pm$0.26\\
\hline\hline
\end{tabular}
\end{table}

\begin{table}[h]
\caption{Summary of the Calar Alto Survey. For each observed exoplanet host star we list the total
exposure time, the measured seeing, the inner and outer detection radii of stellar companions, and
the mass limit of detectable objects.} \label{tab2}
\resizebox{\columnwidth}{!}{\begin{tabular}{lccccc}\hline
star & time & seeing & r$\rm_{in}$  & r$\rm_{out}$ & mass limit\\
& [min] & [arcsec]& [AU] & [AU] & [M$\rm_{Jup}$]\\

\hline

47\,UMa      & 30  & 2.4   & 107 & 946 & 53\\

55\,Cnc      & 30  & 1.0   & 59  & 810 & 37\\

BD$-$10$^{\circ}$31666  & 9   & 1.7   & 369 & 4287 & 72\\

HD\,8574     & 30  & 1.0   & 314 & 2741 & 56\\

HD\,10697    & 30  & 1.2   & 261 & 2042 & 56\\

HD\,37124    & 25  & 1.6   & 239 & 2298 & 62\\

HD\,38529    & 26  & 1.7   & 433 & 2933 & 66\\

HD\,46375    & 26  & 1.4   & 190 & 2138 & 65\\

HD\,50554    & 30  & 1.3   & 199 & 2105 & 63\\

HD\,52265    & 33  & 1.1   & 154 & 1904 & 54\\

HD\,74156    & 32  &  1.5  & 516 & 4256 & 70\\

HD\,80606    & 31  & 1.1   & 321 & 2877 & 66\\

HD\,82943    & 27  & 1.2   & 179 & 1775 & 55\\

HD\,92788    & 30  &  1.7  & 223 & 2089 & 61\\

HD\,106252   & 33  & 1.4   & 251 & 2396 & 63\\

HD\,136118   & 30  &  1.2  & 403 & 3412 & 63\\

HD\,114762   & 23  & 1.5   & 316 & 2674 & 66\\

HD\,209458   & 24  &  1.0  & 235 & 3073 & 59\\
\hline\hline
\end{tabular}}
\end{table}

Because most of the observed exoplanet host stars are nearby relatively bright stars the
integration time had to be reduced to only a few tenth of a second, in order to avoid saturation
effects of the detector. Many of these short exposures were averaged to one frame with a total
integration time of $\sim$\,3~minutes. The telescope was then moved by a few arcsec and the
integration procedure was repeated. With this jitter observing technique the bright infrared sky
background could effectively be subtracted from the images. To correct for the individual pixel
sensitivity all frames were flatfielded using sky flat frames which were taken at the beginning of
the night in twilight. Background subtraction, flatfielding, image registration, shifting and final
averaging of all images were achieved with the data reduction package ESO eclipse
\footnote{ECLIPSE: ESO C Library for an Image Processing Software Environment} (Devillard 2001).

The achieved observational data of all exoplanet host stars observed with MAGIC in the two
observing runs carried out in January and July 2002 are summarized in Table\,\ref{tab2}. We list
the total integration time per target (\textsl{time}) and the average seeing measured in each MAGIC
image (\textsl{seeing}). In general, after 30~min of integration a detection limit (S/N\,=\,3) of
H\,=\,18\,mag was reached. With the given distances of the exoplanet host stars we can convert the
detection limits to the absolute magnitudes of the faintest detectable companions. The mass of
these companions can be approximated with theoretical models, using mass-magnitude relations
therein. With Baraffe et al. (2003) models and an average system age of 5\,Gyr we estimate to be
sensitive to substellar companions down to $\sim$\,60\,M$\rm_{Jup}$. Table\,\ref{tab2} shows the
derived mass limit of any detectable companions as well as the inner and outer detection radii
(\textsl{r$\rm_{in}$}, \textsl{r$\rm_{out}$}) of stellar companions in all MAGIC images. The inner
limit depends on the brightness of the primary and on the seeing. The outer radius is only limited
by the MAGIC field of view. In average additional stellar companions can be detected around all
targets in a range of separations between about 270\,AU and 2500\,AU.

\begin{figure}[h]
\resizebox{\hsize}{!}{\includegraphics[]{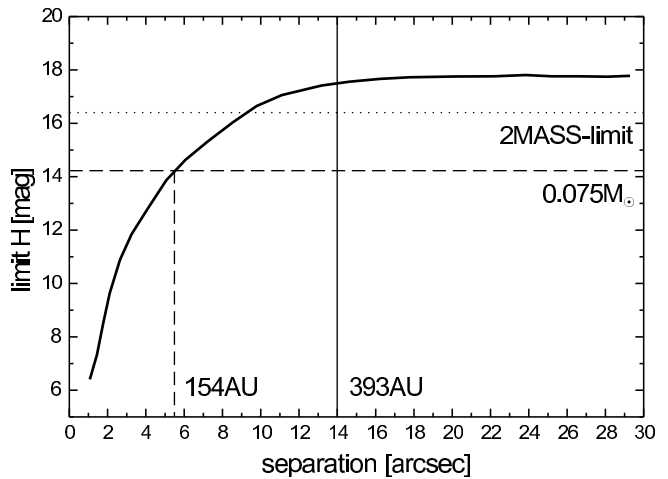}} \caption{The achieved MAGIC detection limit
(S/N\,=\,3) for a range of separations to the exoplanet host star HD\,52265. Beyond 14\,arcsec
(393\,AU), where the noise is dominated by the background, the sensitivity goes down to
H\,=\,17.9\,mag (see solid line). According to Baraffe et al. (2003) models, the limiting magnitude
enables the detection of substellar companions down to 54\,M$\rm_{Jup}$, assuming a system age of
5\,Gyr. Further stellar companions can be ruled out between 154\,AU and 1904\,AU in projected
separation (see dashed lines). The 2MASS detection limit is shown as a dotted line.} \label{limit}
\end{figure}

As a typical example, we plot in Fig.\,{\ref{limit}} the achieved detection limit as a function of
separation to the exoplanet host star HD\,52265. Stellar companions can be detected from
5.5\,arcsec (154\,AU) up to 67.8\,arcsec (1904\,AU). A limiting magnitude of 17.9\,mag is reached
beyond 14\,arcsec (393\,AU). This is about $\sim$\,1.5\,mag deeper than the 2MASS detection limit,
i.e. all objects detected in 2MASS are also visible in our MAGIC images. However close and faint
companions around all these stars are not accessible for our wide field imaging. For example with
AO imaging Patience et al. (2002) could detected a close companion of HD\,114762
(H\,$\sim$\,13.4\,mag) which is located only 3.3\,arcsec north-east of the exoplanet host star.
This object is not visible in our MAGIC image as expected from the derived detection limit.

\begin{figure}[h!]
\resizebox{\hsize}{!}{\includegraphics[]{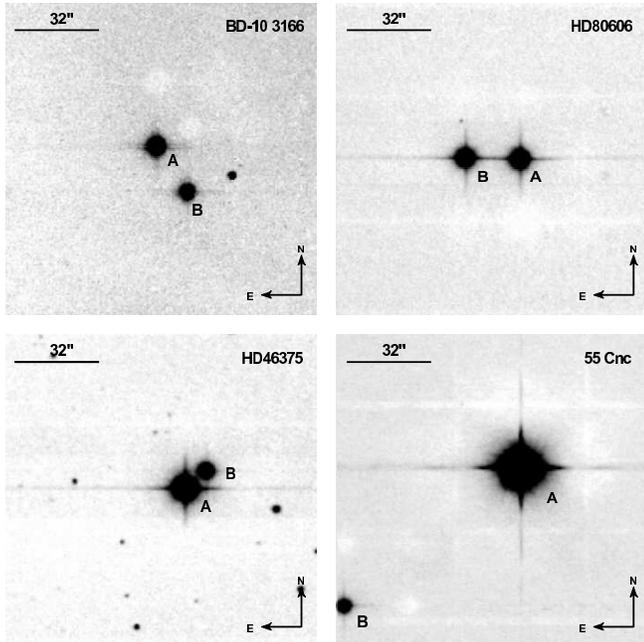}} \caption{The four wide WDS binaries among the
exoplanet host stars observed with MAGIC at the Calar Alto 2.2m telescope. The system
BD$-$10$^{\circ}$3166 turns out to be only a visual double star (see section 3 for details).}
\label{pics}
\end{figure}

Because real companions of the exoplanet host stars are co-moving to their parent stars, i.e. both
objects form a common proper motion pair, they can be distinguished from unrelated slow or
non-moving background stars. By comparing all our MAGIC images with images from 2MASS and
POSS\footnote{Palomar All Sky Survey} I and II, the companionship of all detected objects in the
MAGIC images can be checked. Among the 18 observed exoplanet host stars, no further so far unknown
co-moving companions could be detected around the target stars.

However our sample contains four systems which are listed as binaries in the Washington Visual
Double Star (WDS) catalog (Worley 1997). The MAGIC images of these four systems are shown in
Fig.\ref{pics}. The measured separations and position angles are summarized in Table\,\ref{tab3}
together with the distances of the exoplanet host stars. Thereby the distances of HD\,80606,
55\,Cnc and HD\,46375 can be derived from Hipparcos parallaxes (Perryman et al. 1997).
BD$-$10$^{\circ}$3166 is not listed in the Hipparcos catalogue therefore no accurate parallax is
available for this star. To estimate the distance of BD$-$10$^{\circ}$3166 we use its spectral
type, K0V, derived by Butler et al. (2000), as well as photometric data in different filters
(B\,=\,10.727$\pm$0.081\,mag, V\,=\,10.000$\pm$0.044\,mag\footnote{B,V magnitudes from Kharchenko
(2001)}, J\,=\,8.611$\pm$0.032\,mag, H\,=\,8.300$\pm$0.040\,mag, and
K$\rm{_{S}}$\,=\,8.124$\pm$0.026\,mag\footnote{J,H,K$\rm{_{S}}$ from the 2MASS point source
catalogue}). The apparent photometry of BD$-$10$^{\circ}$3166 is consistent with a K0V dwarf at a
distance of 67$\pm$3\,pc.

\begin{table}[ht]
\caption{The astrometry of the four WDS binary systems, observed with MAGIC. We show the
separations and position angles for all systems, measured in the MAGIC images. In addition we list
the distance of the four exoplanet host stars which are all determined with Hipparcos parallaxes,
expected BD$-$10$^{\circ}$3166 whose distance is derived using the spectral type of the star and
its apparent magnitudes.} \label{tab3} \centering{\begin{tabular}{lccc}\hline
star & distance & separation & PA\\
& [pc] & [arcsec] & [$^{\circ}$]\\
\hline
HD\,80606 & 58 & 20.549$\pm$0.122 &\,\,88.66$\pm$0.20 \\
\hline
HD\,46375 & 34 & 10.352$\pm$0.061&309.95$\pm$0.31 \\
\hline
BD$-$10$^{\circ}$3166 & 67 & 20.891$\pm$0.124 &214.21$\pm$0.29 \\
\hline
55\,Cnc & 13 & 84.715$\pm$0.503 &128.10$\pm$0.30 \\
\hline\hline
\end{tabular}}
\end{table}

The proper motions of both components of the four observed WDS binaries are already given by the
Hipparcos, the USNO-B1.0 (Monet et al. 2003) and the UCAC2 catalogues (Zacharias et al. 2004). Only
the secondary of HD\,46375 is not listed in any of these catalogues. HD\,46375 is not well resolved
in the POSS images but the companion is clearly separated from the bright primary star in the 2MASS
image. By comparing this image with our MAGIC image we can derive the proper motion of HD\,46375\,B
with precision which is limited mainly by the 2MASS astrometric accuracy. We summarize the proper
motions of the four WDS binaries in Table\,\ref{tab4}.

\begin{table}[htb]
\caption{Proper motions of the four WDS binary systems observed with MAGIC. The proper motion of
all primaries as well as three of the four secondaries are already published by Hipparcos (HIP),
USNO-B1.0 and UCAC2 catalogues. The proper motion of HD\,46375\,B was derived by comparing the
2MASS image with our MAGIC image.} \label{tab4}
\resizebox{\columnwidth}{!}{\begin{tabular}{lccc}\hline
star & $\rm\mu_{Ra}$ & $\mu\rm_{Dec}$ & reference\\
& [mas/yr] & [mas/yr]\\
\hline
HD\,80606\,A      & 46.98$\pm$6.32   & 6.92$\pm$3.99    & HIP\\
HD\,80606\,B      & 42.80$\pm$9.32   & 8.26$\pm$5.88    & HIP\\
\hline
HD\,46375\,A      & 114.24$\pm$0.92  & -96.79$\pm$0.73  & HIP\\
HD\,46375\,B      & 134$\pm$45       & -128$\pm$43      & MAGIC-2M\\
\hline
BD$-$10$^{\circ}$3166\,A    & -186.8$\pm$1.6   & -6.2$\pm$1.3     & UCAC2\\
BD$-$10$^{\circ}$3166\,B    & -198.8$\pm$4.4   & -93.0$\pm$4.4    & UCAC2\\
\hline
55\,Cnc\,A        & -485.48$\pm$0.98 & -234.40$\pm$0.78 & HIP\\
55\,Cnc\,B        & -488$\pm$6       & -234$\pm$5       & USNO-B1.0\\
\hline\hline
\end{tabular}}
\end{table}

The 2MASS point source catalogue provides accurate near infrared photometry of the secondaries and
Kharchenko (2001) lists V band magnitudes for HD\,80606\,B and BD$-$10$^{\circ}$3166\,B. The V band
magnitudes of 55\,Cnc\,B and HD\,46375\,B are listed in the WDS catalogue but no photometric
uncertainties are given there (see Table\,\ref{photodata}).

\begin{table}[htb]
\centering{\caption{Photometric data of the secondaries of the four WDS binaries observed with
MAGIC. We list the J, H and K$\rm{_{S}}$ near-infrared magnitudes taken from the 2MASS point source
catalogue. All V band magnitudes with given uncertainties are taken from Kharchenko (2001). The V
magnitudes of HD\,46375\,B and 55\,Cnc\,B are listed in the WDS catalogue but without
uncertainties.} \label{photodata} \resizebox{\columnwidth}{!}{\begin{tabular}{lcccc}\hline
companion & J & H & K$\rm{_{S}}$ & V\\
& [mag] & [mag] & [mag] & [mag]\\
\hline
HD\,80606\,B & 7.798$\pm$0.029 & 7.509$\pm$0.029 & 7.389$\pm$0.021 & 9.090$\pm$0.022\\
\hline
HD\,46375\,B & 8.701$\pm$0.034 & 8.083$\pm$0.053 & 7.843$\pm$0.021 & 11\\
\hline
BD$-$10$^{\circ}$3166\,B & 9.512$\pm$0.023 &  8.965$\pm$0.022& 8.640$\pm$0.021 & 14.437$\pm$0.153\\
\hline
55\,Cnc\,B & 8.560$\pm$0.027 & 7.933$\pm$0.040 & 7.666$\pm$0.023 & 13.16\\
\hline\hline
\end{tabular}}}
\end{table}

\section{Discussion}

We observed 18 exoplanet host stars with MAGIC at Calar Alto. New co-moving companions could not be
detected but 4 binary systems were observed which are already listed in the Washington Visual
Double star catalogue (WDS). The proper motions of the primary and the secondary components of
these binary systems are summarized in Table\,\ref{tab4}. Both components of the WDS binaries
HD\,80606, HD\,46375, and 55\,Cnc share a common proper motion. Therefore the companionship of
these systems is confirmed by astrometry.

The WDS binary BD$-$10$^{\circ}$3166 consists of two high proper motion stars which exhibit proper
motions being similar in right ascension (Ra) but significantly differ in Declination (Dec).
According to the astrometric UCAC2 catalogue (see Table\,\ref{tab4}) the motion of the secondary
relative to the primary is $-$12$\pm$4.7\,mas/yr in Ra but $-$86.8$\pm$4.6\,mas/yr in Dec, i.e. a
total relative motion of 87.6$\pm$4.6\,mas/yr. Comparing our MAGIC image with the 2MASS and the
POSS-I and POSS-II images yields a similar result. We derive a motion of the secondary relative to
the primary component of $-$13$\pm$7\,mas/yr in Ra and $-$79$\pm$7\,mas/yr in Dec. Therefore we can
conclude that this WDS binary is clearly not a common proper motion pair.

\begin{figure}[t!]
\resizebox{\hsize}{!}{\includegraphics[]{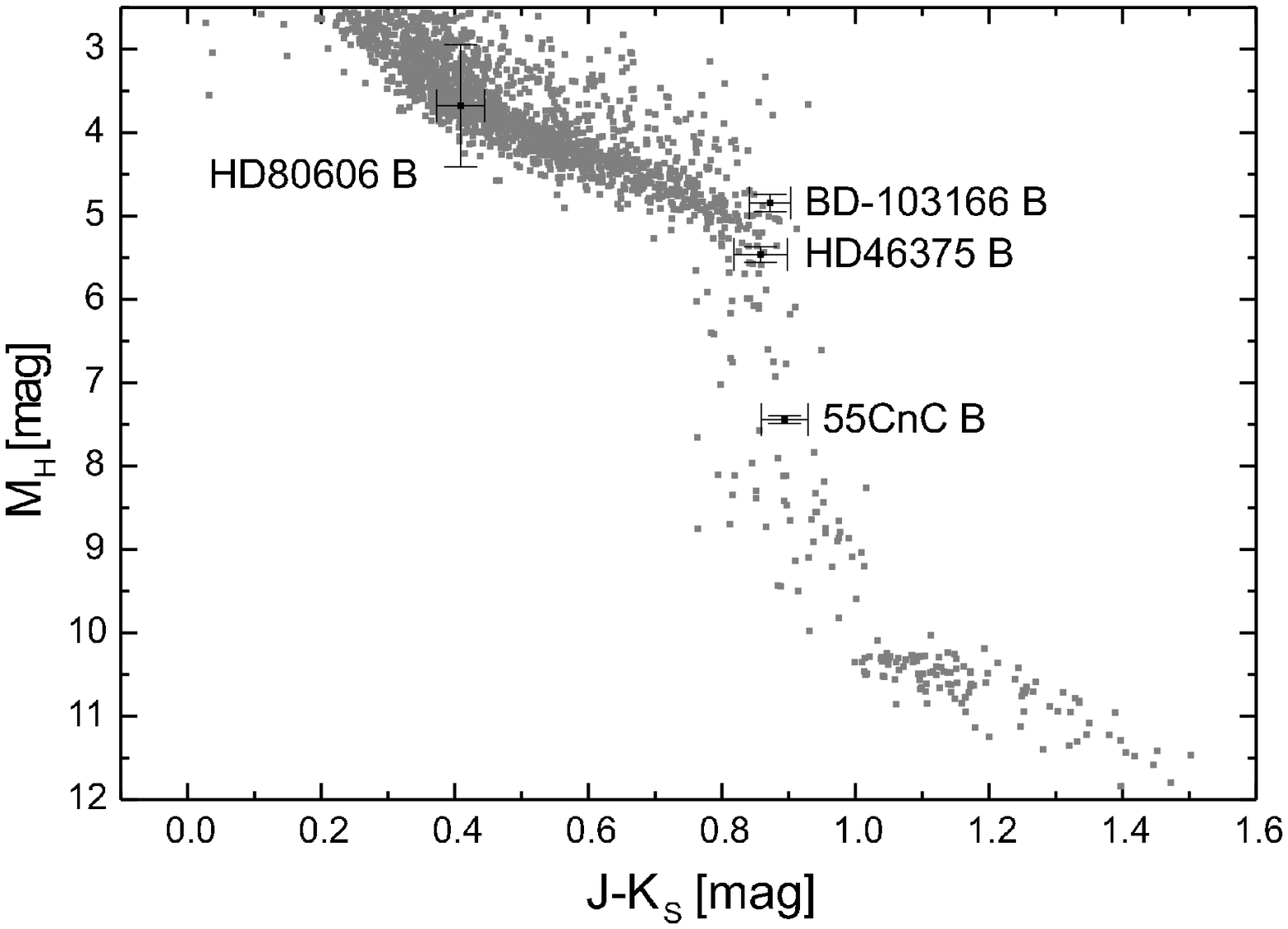}}

\resizebox{\hsize}{!}{\includegraphics[]{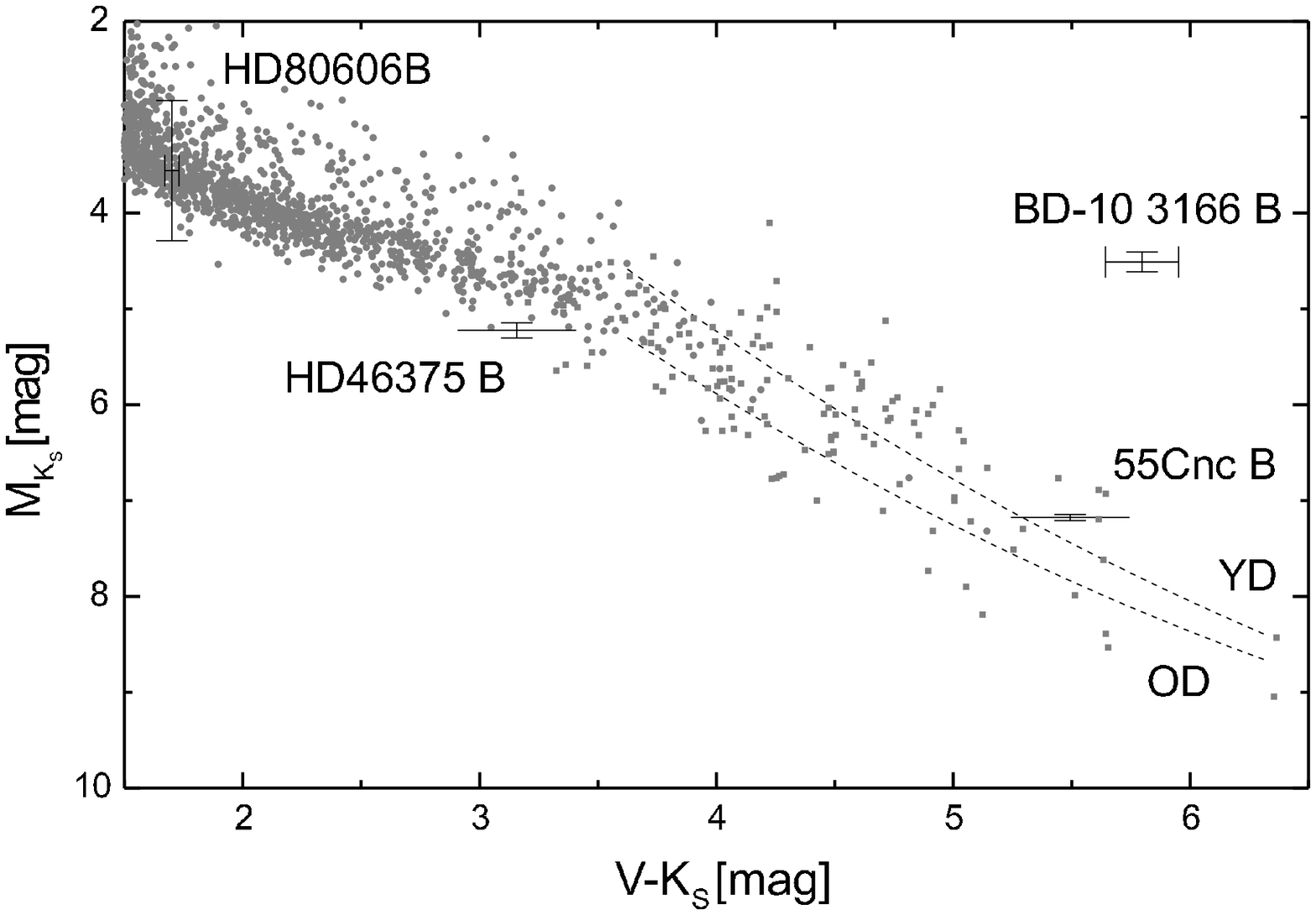}} \caption{The photometric test of
companionship of the four WDS binary systems. We plot the four secondaries in a J-K$\rm{_{S}}$,
M$_{\rm H}$ (top panel) and a V-K$\rm{_{S}}$, M$_{\rm K_{S}}$ diagram (bottom panel). The absolute
magnitudes of the secondary stars are derived with their apparent magnitudes and the distances of
the exoplanet host stars. For comparison we show colors and absolute magnitudes of dwarfs taken
from the Hipparcos catalogue, the Nearby Stars catalogue (Gliese \& Jahrei\ss~1995), and from Cruz
et al. (2003). We select only those dwarfs for which accurate photometric data are available
($\sigma(J-K\rm{_{S}})<0.05$\,mag, $\sigma(M_{\rm H})<0.25$\,mag). In the bottom panel we use the
same Hipparcos comparison dwarfs as in the upper plot (filled circles). In addition we plot young
and old disk dwarfs (filled squares) from Leggett et al. (1992) together with the least-square fits
(dashed lines) for these two dwarf populations, as derived by Leggett et al. (1992). The V band
magnitude of HD\,46375\,B and 55\,Cnc\,B are taken from the WDS catalogue but no V band
uncertainties are published there. This is illustrated with open error bars. The photometry of
BD$-$10$^{\circ}$3166\,B is inconsistent with a dwarf located at the distance of the exoplanet host
star but fully comply with a M4 to M5 dwarf located only at a distance of 13\,pc, i.e. it is just
an unrelated foreground object.} \label{photoir}
\end{figure}

In a next step we can test the companionship of the four WDS binaries with photometry. We derive
the absolute magnitudes of the four secondaries using their apparent magnitudes. Thereby we always
assume that the companions are located at the distances of the exoplanet host stars, as it is
expected for real companions. For all secondaries accurate 2MASS near infrared photometry is
available and Kharchenko (2001) provides V band magnitudes of HD\,80606\,B and
BD$-$10$^{\circ}$3166\,B. The V band magnitudes of 55\,Cnc\,B and HD\,46375\,B (photometric
uncertainties are not available) are listed in the WDS catalogue (see Table\,\ref{photodata}). We
plot all secondaries in a J-K$\rm{_{S}}$, M$_{\rm H}$ diagram (upper panel of Fig.\ref{photoir}),
together with comparison dwarfs from the Hipparcos catalogue, the Nearby Stars catalogue (Gliese \&
Jahrei\ss~1995), and cool dwarfs from Cruz et al. (2003). Only comparison objects with accurate
colors ($\rm\sigma(J-K_{S})<0.05$) and accurate absolute magnitudes ($\rm\sigma(M_{\rm H})<0.25$)
are plotted. Furthermore we plot all secondaries in a V-K$\rm{_{S}}$, M$_{\rm K\rm{_{S}}}$ diagram
(see bottom panel of Fig.\ref{photoir}), using again the same Hipparcos comparison dwarfs as in the
infrared color-magnitude diagram, as well as M dwarfs from Leggett et al. (1992).

\begin{table}[hbt]
\centering{\caption{Summary of all photometric data which are shown in Fig.\,\ref{photoir}. We list
the J-K$\rm{_{S}}$ colors as well as the absolute H and K band magnitudes of the companions derived
from the apparent 2MASS magnitudes and the known distances of the exoplanet host stars. The
V-K$\rm{_{S}}$ color is derived with K$\rm{_{S}}$ from 2MASS and V band magnitudes taken either
from Kharchenko (2001) or from the WDS catalogue (see also Table\,\ref{photodata}).}
\label{photodata2} \resizebox{\columnwidth}{!}{\begin{tabular}{lccccc}\hline
companion & J-K$\rm{_{S}}$ & M$_{\rm H}$ & V-K$\rm{_{S}}$ & M$_{\rm K\rm{_{S}}}$\\
& [mag] & [mag] & [mag]& [mag]& \\
\hline
HD\,80606\,B & 0.409$\pm$0.036 & 3.678$\pm$0.732 & 1.701$\pm$0.03 & 3.558$\pm$0.731\\
\hline
HD\,46375\,B & 0.858$\pm$0.040 & 5.464$\pm$0.094 & 3.157 & 5.224$\pm$0.081\\
\hline
BD$-$10$^{\circ}$3166\,B & 0.872$\pm$0.031 & 4.835$\pm$0.103 & 5.797$\pm$0.154 & 4.510$\pm$0.103\\
\hline
55\,Cnc\,B & 0.894$\pm$0.035 & 7.443$\pm$0.046 & 5.494 & 7.176$\pm$0.033\\
\hline\hline
\end{tabular}}}
\end{table}

According to the intrinsic colors for dwarfs and giants published by Tokunaga (2000) the
V-K$\rm{_{S}}$ and J-K$\rm{_{S}}$ colors of BD$-$10$^{\circ}$3166\,B are both fully consistent with
a M4 to M5 dwarf. Nevertheless if we assume that this object is indeed located at the distance of
the exoplanet host star the derived absolute K$\rm{_{S}}$ band magnitude is about 3.5 mag brighter
than comparison dwarfs with the same V-K$\rm{_{S}}$ color (see the bottom panel of
Fig.\,\ref{photoir}), i.e. the distance of this object is overestimated by a factor of 5. The
apparent infrared and V band photometry of BD$-$10$^{\circ}$3166\,B is consistent with a M4 to M5
dwarf located at a distance of about 13\,pc. Therefore we can conclude that the WDS binary
BD$-$10$^{\circ}$3166 is only a visual pair of stars, consisting of a K0 dwarf in the background at
a distance of 67\,pc and an unrelated foreground M dwarf at a distance of about 13\,pc.

As it is illustrated in the color-magnitude diagrams shown in Fig.\,\ref{photoir} the photometry of
the secondaries of the common proper motion pairs HD\,80606, HD\,46375 and 55\,Cnc is consistent
with dwarfs located at the distances of the exoplanet host stars, hence the companionship of the
these three WDS binaries is confirmed by both astrometry and photometry.

The masses of the three secondaries HD\,80606\,B, 55\,Cnc\,B, and HD\,46375\,B can be derived by
converting their absolute infrared magnitudes to masses, using the evolutionary Baraffe et al.
(1998) models. Thereby we always assume a system age of 5\,Gyr, which is a good estimate for most
of the exoplanet host stars. It is important to mention that, the age uncertainty of the binary
systems in the order of a few Gyr is not important here because for system ages between 1 and
10\,Gyr the infrared magnitudes of stars with masses below one solar mass are not strongly age
dependant. The derived companion masses range from $\rm{0.9\,M_{\odot}}$ to $\rm{0.27\,M_{\odot}}$,
with mass ratios (M$\rm_{c}/M\rm_{p}$) between 0.3 to 0.87.

\begin{table}[ht]
\caption{Summary of derived properties of the binary systems. For all three binaries imaged with
MAGIC we show their projected separations, derived from the angular separation and the distance of
the system, and the masses of primary (M$\rm_{p}$) and secondary M$\rm_{c}$. For each system we
show the critical semi-major axis for a circular and an eccentric (e\,=\,0.8) binary orbit,
respectively. The primary masses are all derived by Santos et al. (2004).} \label{tab5}
\resizebox{\columnwidth}{!}{
\begin{tabular}{lccccccc}\hline
companion & sep. & M$\rm_{p}$ & M$\rm_{c}$ & a$_{\rm{c, e\,=\,0.0}}$ & a$_{\rm{c, e\,=\,0.8}}$\\
& [AU] & [M$\rm_{\sun}$] & [M$\rm_{\sun}$] & [AU] & [AU]\\
\hline
HD\,80606\,B      &     1200$\pm$404   & 1.04 & 0.904$\pm$0.147 & 345 & 45\\
\hline
HD\,46375\,B      &     346$\pm$13    & 0.82 & 0.576$\pm$0.013 & 106 & 14\\
\hline
55\,Cnc\,B        &     1062$\pm$13   & 0.87 & 0.265$\pm$0.006 & 399 & 49\\
\hline\hline
\end{tabular}}
\end{table}

Finally with the measured angular separations and known distances of the binary systems we can
derive their projected separations. Because neither the orientation of the binary orbit to the line
of sight nor the position of both stars on their orbits around their common barycenter is known, we
always use the observed projected separation also as estimate of the binary semi-mayor axis. The
separation of the three binaries range from 350 up to 1200\,AU.

According to Holman \& Wiegert (1998) objects, e.g. additional companions, located in a binary
system are only longterm stable if their semi-major axes do not exceed a critical value, the
critical semi-major axis ($\rm{a_{c}}$), which depends on the binary semi-major axis, its
eccentricity as well as its mass-ratio. The critical semi-major axes of possible additional
companions in the three binary systems are listed in Table\,\ref{tab5}. They range from about
400\,AU for an assumed circular binary orbit down to only 10\,AU for an eccentric binary system,
respectively. Our observations didn't reveal any further, wide co-moving companions in these
systems, as expected from the Holman \& Wiegert (1998) stability criteria.

Eggenberger et al. (2004) have recently compared the statistical characteristics of planets in
binary systems with those orbiting single stars. They found that the distribution of the masses of
binary-star planets with periods shorter than 40 days is approximately flat, whereas single-star
planets all exhibit masses less than $\rm{2M_{Jup}}$. Furthermore, Eggenberger et al. (2004) found
that all known close binary-star planets have almost circular orbits (e$<$0.05), while eccentric
orbits are only detected among single-star planets.

However, these statistical differences are based only on a small number of known binary-star
planets, and therefore the significance of these differences is still not clear. Note that the
whole sample of the exoplanet host stars has not been systematically surveyed so far for close or
wide companions. Only systematic search programs for companions can clarify the multiplicity status
of the stars in the sample.

In our study we have already shown that there are indeed several exoplanet host stars considered as
single stars in the published statistical analyses which emerge as binary systems (see e.g.
Mugrauer et al. 2005b). A further example of these former unknown or unconfirmed binary systems
among the exoplanet host stars is HD\,46375, whose binary nature was confirmed here for the first
time with astrometry as well as photometry. The exoplanet in that binary system was detected by
Marcy et al. (2000). It is a hot Jupiter (Msin(i)\,=\,0.249\,$\rm{M_{Jup}}$) which revolves its
parent star in only 3.024 days on an almost circular (e\,=\,0.04) orbit, typical for such a short
period binary-star planet.

The Calar Alto imaging survey was just the beginning of our multiplicity study of the exoplanet
host stars, using a 2\,m class telescope. Only a relative small number of exoplanet host stars was
observed. We expanded our search for additional wide companions of exoplanet host stars, using
larger mirrors to be sensitive to fainter companions. On the northern sky we use the UKIRT on
Hawaii and southern targets are observed with the NTT on La Silla, and since begin of 2005 also
with the VLT on Paranal. The results of these surveys finally will clarify the multiplicity status
of most of the exoplanet host stars and will verify the significance of the reported statistical
differences between single-star and binary-star planets.

\acknowledgements{We would like to thank the technical staff of Calar Alto observatory for
assistance in carrying out all observations. Furthermore this publication makes use of data
products from the Two Micron All Sky Survey, which is a joint project of the University of
Massachusetts and the Infrared Processing and Analysis Center/California Institute of Technology,
funded by the National Aeronautics and Space Administration and the National Science Foundation, as
well as the SIMBAD and VIZIER databases, operated at CDS, Strasbourg, France. MF has received
support from the Deutsches Zentrum f\"ur Luft- und Raumfahrt (DLR), F\"orderkennzeichen 50 OR 0401,
and from the Spanish grant AYA2004-05395.}

\end{document}